\documentclass{aa}
\usepackage{graphicx,amssymb,psfig,epsfig}

\begin{document}
\bibliographystyle{natbib}

\title{Evaluating Gaia performances on eclipsing binaries.}
\subtitle{IV. Orbits and stellar parameters for SV~Cam, BS~Dra and HP~Dra}

\author{
E.F. Milone\inst{1}
\and   U. Munari\inst{2,3}
\and   P.M. Marrese\inst{2,3}
\and   M.D. Williams\inst{1}
\and   T. Zwitter\inst{4}
\and   J. Kallrath\inst{5,6}
\and   T. Tomov\inst{7}
       }
\offprints{U.Munari}

\institute {
Physics and Astronomy Department, University of Calgary, Calgary T2N 1N4, Canada
\and 
Osservatorio Astronomico di Padova, Sede di Asiago, I-36012 Asiago (VI), Italy
\and
Dipartimento di Astronomia dell'Universit\`a di Padova, Osservatorio Astrofisico, I-36012 Asiago (VI), Italy
\and 
University of Ljubljana, Department of Physics, Jadranska 19, 1000 Ljubljana, Slovenia 
\and
BASF-AG, Scientific Computing (GVC/S-B009), D67056 Ludwigshafen, Germany
\and
Department of Astronomy, University of Florida, Gainesville, FL, USA
\and
Centre for Astronomy, Nicholaus Copernicus University, ul. Gagarina 11, 87-100 Torun, Poland
}
\date{Received date..............; accepted date................}

\abstract{ 
This is the fourth in a series of papers that aim both to provide reasonable
orbits for a number of eclipsing binaries and to evaluate the expected
performance of Gaia of these objects and the accuracy that is achievable in
the determination of such fundamental stellar parameters as mass and radius.
In this paper, we attempt to derive the orbits and physical parameters for
three eclipsing binaries in the mid-F to mid-G spectral range.  As for
previous papers, only the $H_{\rm P}$, $V_{\rm T}$, $B_{\rm T}$ photometry
from the Hipparcos/Tycho mission and ground-based radial velocities from
spectroscopy in the region 8480$-$8740 \AA\ are used in the analyses.  These
data sets simulate the photometric and spectroscopic data that are expected
to be obtained by Gaia, the approved ESA Cornerstone mission to be launched
in 2011.  The systems targeted in this paper are SV~Cam, BS~Dra and
HP~Dra. SV~Cam and BS~Dra have been studied previously, allowing 
comparisons of the derived parameters with those from full scale and devoted
ground-based investigations.  HP~Dra has no published orbital solution.
SV~Cam has a $\beta$ Lyrae type light curve and the others have Algol-like
light curves.  SV~Cam has the complication of light curve anomalies, usually
attributed to spots; BS~Dra has non-solar metallicity, and HP~Dra appears to
have a small eccentricity and a sizeable time derivative in the argument of
the periastron. Thus all three provide interesting and different test cases.

\keywords{surveys:Gaia -- stars: fundamental parameters -- binaries:eclipsing -- 
binaries:spectroscopic}
}

\maketitle

%%%%%%%%%%%%%%%%%%%%%%%%%%%%%%%%%%%%%%%%%%%%%%%%%%%%%%%%%%%%%%%%%%%%%%%%%%%%%%%%%%%%%
\section{Introduction}
%%%%%%%%%%%%%%%%%%%%%%%%%%%%%%%%%%%%%%%%%%%%%%%%%%%%%%%%%%%%%%%%%%%%%%%%%%%%%%%%%%%%%

Gaia is a Cornerstone mission re-approved by ESA in May 2002. It is intended
to perform three important tasks: ({\sl a}) micro-arcsec astrometry, ({\sl
b}) photometry in 4 broad and 11 intermediate passbands and ({\sl c})
R~$\sim$~11\,500 resolution spectroscopy in the $\lambda\lambda$8480$-$8740
\AA\ wavelength region. The completeness limit for both astrometry and
photometry is anticipated to be $V$~$\sim$~20~mag.  Each target star is
expected to be measured around a hundred times during the five year mission
life-time, in an operational mode similar to that of {\sl Hipparcos}. The
astrophysical goals and technical specifications of the mission can be
found, among others, in ESA's {\sl Concept and Technology Study} (ESA
SP-2000-4), Gilmore et al. (1998), Perryman et al. (2001) and in the
proceedings of Gaia conferences edited by Strai\v{z}ys (1999), Bienaym\'{e}
\& Turon (2002), Vansevi\'{c}ius et al. (2002), Munari (2003) and Turon \&
Perryman (2004).

%%%%%%%%%%%%%%%%%%%%%%%%%%%%%%%%%%%%%%%%%%%%%%%%%%%%%%%%%%%%%%%%%%%%%%%%%%%%%%%%%
% TABLE 1
%%%%%%%%%%%%%%%%%%%%%%%%%%%%%%%%%%%%%%%%%%%%%%%%%%%%%%%%%%%%%%%%%%%%%%%%%%%%%%%%%
\begin{table*}[!t]
\tabcolsep 0.08truecm
\caption{Eclipsing binary targets. Data from the Hipparcos Catalogue. 
         $B_{\rm T}$ and $V_{\rm T}$ are Tycho mean and $H_{\rm P}$ 
         are median magnitude values.}
\begin{center}
\begin{tabular}{lcccccccccrr} \hline
&&&&&&&&&&&\\
Names&&Sp.&$H_{\rm P}$&$B_{\rm T}$&$V_{\rm T}$&$\alpha_{\rm J1991.25}$&$\delta_{\rm J1991.25}$                &parallax&\multicolumn{1}{c}{$\mu^{\ast}_\alpha$}&\multicolumn{1}{c}{$\mu_\delta$}   \\ 
     &&   &           &           &           & (h m s)               &($\degr$~$^{\prime}$~$^{\prime\prime}$)& (mas)  &(mas~yr$^{-1}$)    &(mas~yr$^{-1}$)\\
&&&&&&&&&&&\\ \hline
&&&&&&&&&&&\\
SV Cam & HIP 32015 & G5 & 9.4377 & 10.204 & 9.411 & 06 41 18.89 & $+82$ 16 03.8 & 11.77$\pm$1.07 & $+$41.58$\pm$0.95   & $-$152.91$\pm$1.17 \\
BS Dra & HIP 98118 & F5 & 9.2320 &  9.638 & 9.183 & 19 56 28.79 & $+73$ 36 57.6 & ~4.80$\pm$0.74 & $-$7.95$\pm$0.68    & $-$5.06$\pm$0.72 \\
HP Dra & HIP 92835 & G5 & 8.0888 &  8.679 & 8.032 & 18 54 53.46 & $+51$ 18 29.1 & 12.45$\pm$0.72 & $+$23.35$\pm$0.77   & $+$83.40$\pm$0.90 \\
&&&&&&&&&&&\\
\hline
\end{tabular}
\end{center}
\end{table*}
%%%%%%%%%%%%%%%%%%%%%%%%%%%%%%%%%%%%%%%%%%%%%%%%%%%%%%%%%%%%%%%%%%%%%%%%%%%%%%%%%
% FIGURE 1
%%%%%%%%%%%%%%%%%%%%%%%%%%%%%%%%%%%%%%%%%%%%%%%%%%%%%%%%%%%%%%%%%%%%%%%%%%%%%%%%%
\begin{figure*}[t]
\centerline{\psfig{file=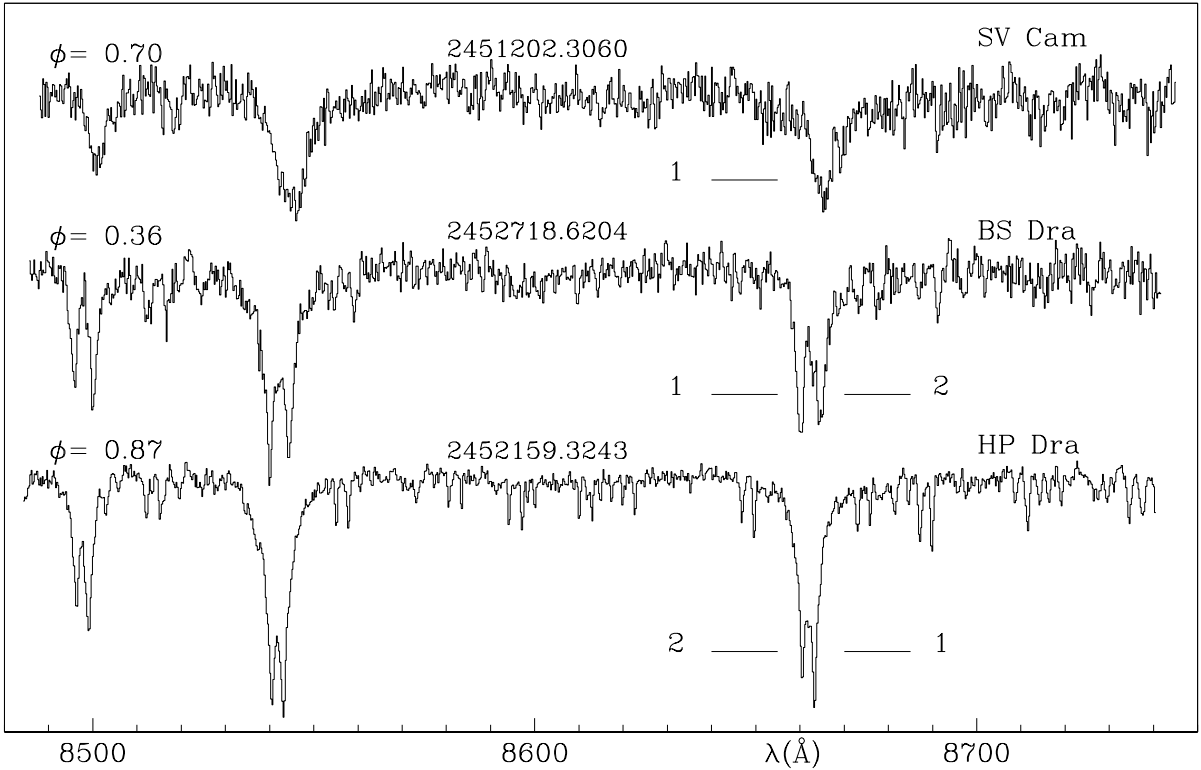,angle=270,width=16cm}}
\caption[]{Sample of the spectra of each of the variables discussed here. Note the lack 
           of visible features of a secondary component in the SV~Cam spectrum contrary to some 
           controversial claims in literature. The spectra are shifted to heliocentric corrected 
           wavelengths.}
\end{figure*}
%%%%%%%%%%%%%%%%%%%%%%%%%%%%%%%%%%%%%%%%%%%%%%%%%%%%%%%%%%%%%%%%%%%%%%%%%%%%%%%%%

In Paper~I of this series, Munari et al. (2001), we began the process of
evaluating the expected performance by Gaia in dealing with eclipsing
binaries and the accuracy with which fundamental stellar parameters such as
masses, radii and temperatures can be determined.  The number of
eclipsing binaries to be discovered by Gaia is expected to be $\sim$10$^6$,
of which $\sim$~10$^5$ are likely to be double-lined. This is orders of
magnitude larger than all the SB2 eclipsing binaries so far investigated
from ground-based observations (e.g. Andersen 1991, 2002), with the addition
of stunningly precise astrometric parallaxes to be used in close-loop iterative
refinement of parameters involved in the orbital modeling. Furthermore,
Gaia's discoveries will be a boon to ground-based observatories by providing
opportunities to follow up discoveries with synoptic, multi-wavelength
campaigns for many decades thereafter.  It is therefore of great interest to
test Gaia's capabilities for eclipsing binary investigations, with the aim of
contributing to the fine tuning of mission planning and preparing analysis
strategies for the massive data flow that will follow.

%%%%%%%%%%%%%%%%%%%%%%%%%%%%%%%%%%%%%%%%%%%%%%%%%%%%%%%%%%%%%%%%%%%%%%%%%%%%%%%%%
% TABLE 2
%%%%%%%%%%%%%%%%%%%%%%%%%%%%%%%%%%%%%%%%%%%%%%%%%%%%%%%%%%%%%%%%%%%%%%%%%%%%%%%%%
\begin{table}[!b]
\tabcolsep 0.08truecm
\caption{Summary of Hipparcos ($H_{\rm P}$) and Tycho ($B_{\rm T}$, $V_{\rm T}$) 
         photometric data and ground based radial velocity observations: number 
         of data, mean S/N and mean standard error (magnitudes for photometry, km~sec$^{-1}$
         for radial velocities) for each of the three program stars.}
\begin{tabular}{lccccccccccc} \hline 
  && \multicolumn{2}{c}{\sl Hip}&& \multicolumn{3}{c}{\sl Tycho}&&\multicolumn{3}{c}{\sl RV}\\ 
  \cline{3-4} \cline{6-8} \cline{10-12} 
  \multicolumn{11}{c}{}\\
        && N    &$\sigma$($H_{\rm P}$) && N   &$\sigma$($B_{\rm T}$)&$\sigma$($V_{\rm T}$)&& N  &S/N  &$\sigma$(RV)\\
SV Cam  && 113  & 0.021                && 167 & 0.34                & 0.31                && 35 & 40  & 10         \\
BS Dra  && ~88  & 0.016                && 143 & 0.20                & 0.19                && 27 & 36  & ~6         \\  
HP Dra  && ~81  & 0.012                && 122 & 0.11                & 0.09                && 29 & 62  & ~3         \\
\hline
\end{tabular}
\end{table}
%%%%%%%%%%%%%%%%%%%%%%%%%%%%%%%%%%%%%%%%%%%%%%%%%%%%%%%%%%%%%%%%%%%%%%%%%%%%%%%%

Paper~I outlined the framework of the project and adopted methodologies, and
the reader is referred to it (and references therein) for further details. 
In short, Hipparcos/Tycho photometry is adopted as a fair representation of
typical Gaia photometric data, and devoted ground-based observations
obtained with the Asiago 1.82m telescope are used to simulate expected Gaia
spectroscopic data. The latter are obtained in the same Gaia wavelength
region ($\lambda\lambda$8480$-$8740 \AA), at a higher resolution
(R=20\,000) than is planned for the satellite (R=11\,500) to compensate
for the lower number of recorded radial velocity points ($\sim$30 vs. the
$\sim$100 expected from Gaia). Gaia data will obviously largely surpass the
Hipparcos/Tycho photometry adopted here in terms of number of epochs, number
and diagnostic capabilities of the photometric bands, and therefore even
better performance than we have found here can be expected from the satellite.

%%%%%%%%%%%%%%%%%%%%%%%%%%%%%%%%%%%%%%%%%%%%%%%%%%%%%%%%%%%%%%%%%%%%%%%%%%%%%%%%%
% TABLE 3
%%%%%%%%%%%%%%%%%%%%%%%%%%%%%%%%%%%%%%%%%%%%%%%%%%%%%%%%%%%%%%%%%%%%%%%%%%%%%%%%%
\begin{table*}[!t]
\tabcolsep 0.08truecm
\centerline{\psfig{file=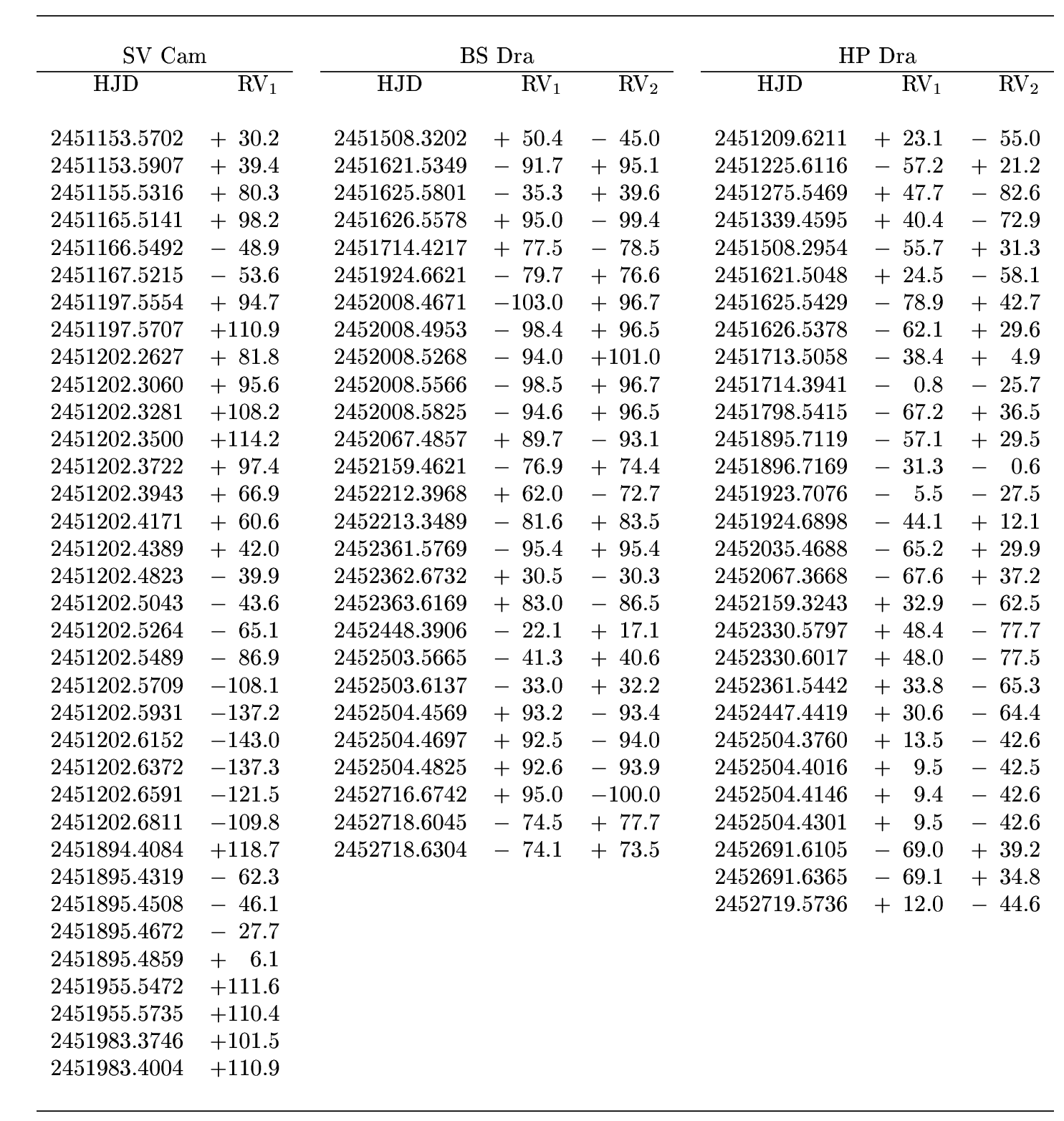,width=16.0cm}}
\caption[]{Journal of radial velocity data. The columns give the Heliocentric 
Julian Day and the radial velocities in km~sec$^{-1}$ for each of the three
systems discussed in this paper.}
\end{table*}
%%%%%%%%%%%%%%%%%%%%%%%%%%%%%%%%%%%%%%%%%%%%%%%%%%%%%%%%%%%%%%%%%%%%%%%%%%%%%%%%%

%%%%%%%%%%%%%%%%%%%%%%%%%%%%%%%%%%%%%%%%%%%%%%%%%%%%%%%%%%%%%%%%%%%%%%%%%%%%%%%%%
\section{The targets}
%%%%%%%%%%%%%%%%%%%%%%%%%%%%%%%%%%%%%%%%%%%%%%%%%%%%%%%%%%%%%%%%%%%%%%%%%%%%%%%%%
As in the previous papers in the series (Munari et al. 2001, Paper I, Zwitter
et al. 2003, Paper II and Marrese et al. 2004, Paper III), we have selected
both newly discovered eclipsing binaries (i.e., those lacking spectroscopic
and photometric orbit solutions in the literature), as well as binaries with
already published orbital solutions (although not with data in the Gaia spectral
range) to permit a comparison. In our analysis, to properly simulate Gaia,
we have ignored all pre-existing ground-based data, our solutions resting
exclusively on Hipparcos/Tycho photometry and Gaia spectroscopy as collected
in Asiago. Similarly, we have intentionally avoided selecting initial
parameters for our modeling trials from published solutions. The three
targets for this fourth paper in the series are SV~Cam, BS~Dra and HP~Dra.
The first one is still a controversial object in spite of hundreds of papers
devoted to it, while some orbital modeling has been already attempted for the second.
The third one is a relatively unstudied eclipsing system. Some basic quantities are given
in Table~1.

{\it SV~Cam}.  This is reported in literature as a mid-F to mid-G system
with 0.6 day period and a light-curve of the $\beta$ Lyrae type.  There is a
long history of observations and analyses of this system, with extensive
photometry (e.g., Kjurkchieva, Marchev \& Og{\l}oza 2000; Patk\'{o}s 1982;
Zeilik et al. 1988), and spectroscopy (e.g., Hempelmann et al. 1997;
Kjurkchieva, Marchev \& Zola 2002; Lehmann, Hempelmann \& Wolter 2002;
\"{O}zeren et al. 2001; Pojma\'{n}ski 1998; Popper 1996; Rainger, Hilditch
\& Edwin 1991 and Rucinski et al. 2002).  Most authors have interpreted
asymmetries in the light curves and the variation in them as due to spots. 
With the inclusion in models of high latitude, long-lasting spots, it is not
surprising that there is substantial disagreement about the spectral type of
the hotter component, with estimates ranging from F5 to G8 (Popper 1996).
The difference in eclipse depths shows that the secondary star, as we refer to 
the component eclipsed at secondary minimum, is much fainter
than the primary (cf. Figure~2) and therefore it is not a surprise that 
it is not easily detectable and measurable in our spectra.  
A similar conclusion was reached by Popper (1996), based on the
photometry of Zeilik et al. (1988) and by Patk\'{o}s and Hempelmann (1994). 
Consequently we include in the analysis only the radial velocity curve of the
hotter, more luminous component. This means that, from our data, the mass ratio
is not independently determined by spectroscopic means, and its determination
must depend on the curvature of the maxima in the light curves, which are
relatively noisy. This system thus becomes an interesting test case for Gaia
analyses, both because of activity and because of the low luminosity of the
secondary.

{\it BS~Dra}. This system is a partially eclipsing Algol system with similar
components and a period of $\sim$3.4 days. The Gaia-like spectra reveal two
stars of nearly identical spectral types and luminosities.  A comparison with
the spectral atlases by Munari \& Tomasella (1999), Marrese et al. (2003)
and Zwitter et al. (2004) obtained in the same wavelength region at the same
(R=20\,000) resolution, indicates that the the spectral type is $\sim$~F3V and
a metallicity lower than solar ([Fe/H]$\sim-$0.4). Previous studies were
carried out by Popper (1971), G{\"{u}}d{\"{u}}r et al. (1979) and by Russo et
al. (1981).

{\it HP~Dra}.  This system was discovered as an eclipsing binary by
Hipparcos, but the reported period (6.67 days) is incorrect.  We find from
our spectroscopic data that this Algolid system has a period of
$\sim$~10.761 days. A similar value has been found also by
Kurpinska-Winiarska et al. (2000). A comparison with the above cited
spectral atlases suggests a spectral type of $\sim$~F9V and a solar
metallicity. The lack of data at minimum light in the Hipparcos and Tycho
data sets proved to be a severe problem in modeling this system, as
discussed below.

%%%%%%%%%%%%%%%%%%%%%%%%%%%%%%%%%%%%%%%%%%%%%%%%%%%%%%%%%%%%%%%%%%%%%%%%%%%%%%%
\section{The data}
%%%%%%%%%%%%%%%%%%%%%%%%%%%%%%%%%%%%%%%%%%%%%%%%%%%%%%%%%%%%%%%%%%%%%%%%%%%%%%%

%%%%%%%%%%%%%%%%%%%%%%%%%%%%%%%%%%%%%%%%%%%%%%%%%%%%%%%%%%%%%%%%%%%%%%%%%%%%%%%
\subsection{Spectroscopic observation and radial velocities}
%%%%%%%%%%%%%%%%%%%%%%%%%%%%%%%%%%%%%%%%%%%%%%%%%%%%%%%%%%%%%%%%%%%%%%%%%%%%%%%

The Gaia simulating spectroscopic observations were obtained with the Echelle and CCD
spectrograph on the 1.82 m telescope operated by Osservatorio Astronomico di
Padova atop Mt. Ekar (Asiago), the same facility and instrument as used for the
targets of the previous papers in this series. The same wavelength range
(8480$-$8740 \AA), reciprocal linear dispersion (0.25~\AA/pix), and
resolution (R~=~$\lambda/\Delta\lambda$~=~20\,000) as for the data reported
in Papers~I, II, and III similarly apply.  Further details of the
spectrograph and the spectroscopy can be found in these earlier papers. The
diary of observations of the stars discussed in this paper appear in
Table~3.  Sample spectra in the Ca~II triplet region for each of the three
systems are shown in Fig.~1. The number of spectroscopic observations and
their accuracy are summarized in Table~2.

%%%%%%%%%%%%%%%%%%%%%%%%%%%%%%%%%
\subsection{Hipparcos/Tycho photometry}
%%%%%%%%%%%%%%%%%%%%%%%%%%%%%%%%
Hipparcos and Tycho epoch photometry was retrieved from CDS. The conversion
of Hipparcos and Tycho photometric quantities to the Johnson-Cousins system,
follows, as in previous papers, the transformation equations provided in the
Hipparcos Catalogue (ESA 1997):
\begin{equation} 
V_{J}~=~V_{T}~-~0.090~\times~(B-V)_{T}\mbox{,} 
\end{equation}
\begin{equation}
(B-V)_{J}~=~0.850~\times~(B-V)_{T} 
\end{equation}
Details on the number of observations and their accuracy are given in Table~2.
The relationships among ($B-V)$, spectral type, temperature, and bolometric 
correction were taken from Popper (1980).

%%%%%%%%%%%%%%%%%%%%%%%%%%%%%%%%%%%%%%%%%%%%%%%%%%%%%%%%%%%%%%%%%%%%%%%%%
\section{The modeling software}
%%%%%%%%%%%%%%%%%%%%%%%%%%%%%%%%%%%%%%%%%%%%%%%%%%%%%%%%%%%%%%%%%%%%%%%%%

We began by using the Wilson-Devinney base program (Wilson 1998), with
modifications noted by Kallrath et al. (1998) and Kallrath \& Milone (1999).
Since Paper~I, the University of Calgary version has further evolved to
wd98k93h, a Unix version which not only makes use of Kurucz stellar
atmosphere models applied to each grid element, but also has sufficiently
numerous grid elements that transit eclipses by objects as small as a
Saturn-sized planet can be modeled.  This version of wd98k93 has been
incorporated into the WD2002 package that runs under Linux or Microsoft 
Windows. The WD2002 package permits self-iterated, damped least squares 
differential corrections, as did its predecessors, and now permits looping 
over a range of a single parameter, such as $q$ or $i$, to improve optimization 
in highly correlated cases.

The program requires tables of the ratio of Kurucz atmospheres to black-body
fluxes in each of the $H_{\rm P}$, $B_{\rm T}$ and $V_{\rm T}$ passbands as
well as in a square passband centered on the Ca~II triplet region, used in
connection with the radial velocity curves. In addition the limb-darkening
coefficients for these passbands are needed. The former were provided by C.R.
Stagg and MDW, and the latter by W. Van Hamme together with
the correct limb darkening coefficients for appropriate effective temperatures,
surface gravities and metallicities. Additionally a two-way coefficient 
interpolator was provided by D. Terrell.

The usual operation begins with reasonable guesses for the initial
parameters based primarily on the spectrosccopic data and secondarily
on the colour indices of the Tycho observations. With WD2002, we try to ascertain
the correct global minimum region of parameter space, by means of simplex 
iterations which explore all the parameter space. Then final parameters are 
found by iterated damped least-squares DC runs. When convergence is achieved, 
the process is concluded. The uncertainties in the orbital parameters quoted
in Table~4 come from final modeling check runs carried out with wd98k93h. 

%%%%%%%%%%%%%%%%%%%%%%%%%%%%%%%%%%%%%%%%%%%%%%%%%%%%%%%%%%%%%%%%%%%%%%%%%%%%%%%%%
% TABLE 4
%%%%%%%%%%%%%%%%%%%%%%%%%%%%%%%%%%%%%%%%%%%%%%%%%%%%%%%%%%%%%%%%%%%%%%%%%%%%%%%%%
\begin{table*}[!t]
   \caption[]{Modeling solutions. The uncertainties are formal mean
   standard errors of the solution. The last four rows give the r.m.s of the   
   observed points from the derived orbital solution. The estimation of error 
   in the adopted temperature for the primaries is $\sim$150 K.}
   \centerline{\psfig{file=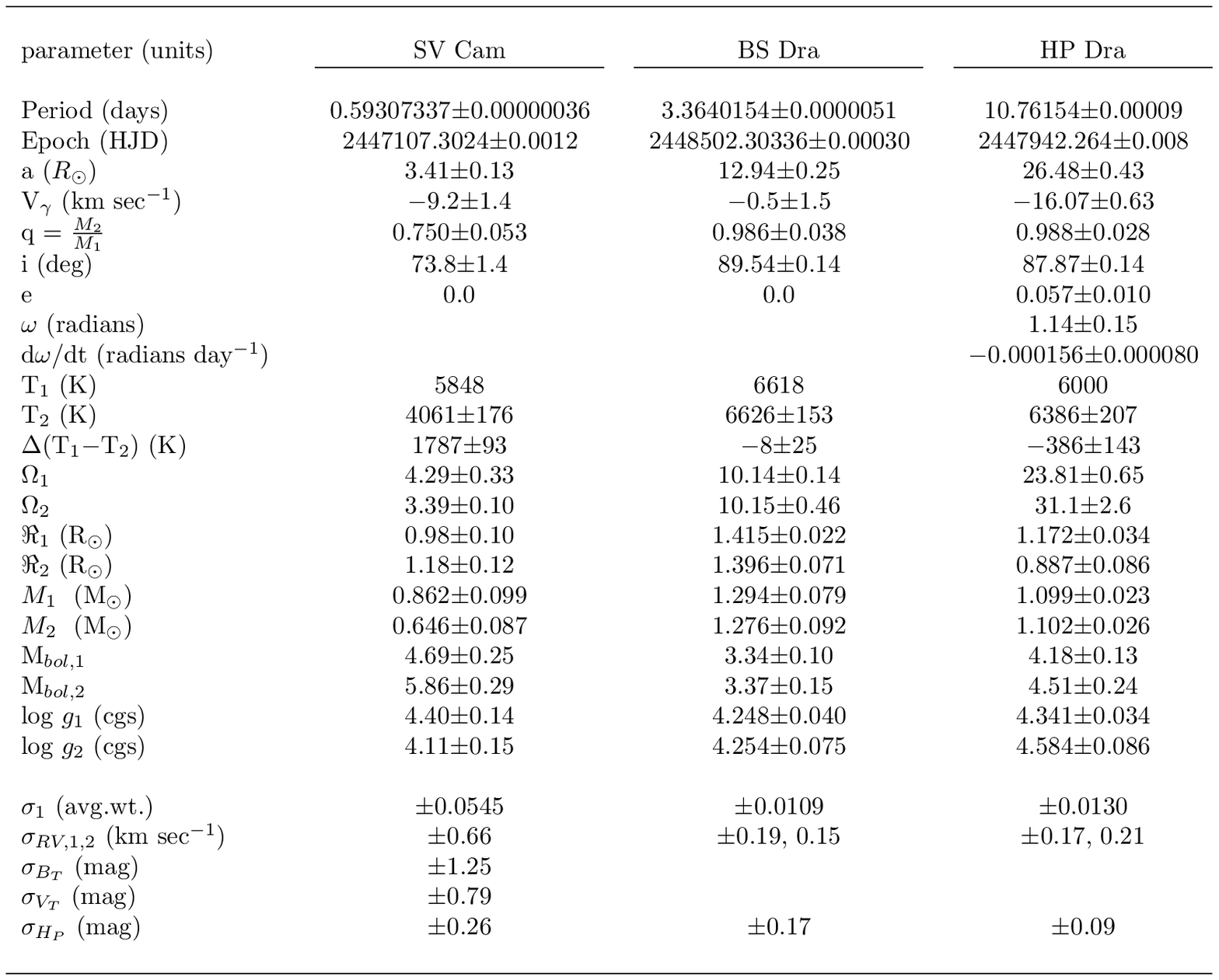,width=15cm}}
   \end{table*}
%%%%%%%%%%%%%%%%%%%%%%%%%%%%%%%%%%%%%%%%%%%%%%%%%%%%%%%%%%%%%%%%%%%%%%%%%%%%%%%%%
%%%%%%%%%%%%%%%%%%%%%%%%%%%%%%
\section{Analyses}
%%%%%%%%%%%%%%%%%%%%%%%%%%%%%%
%%%%%%%%%%%%%%%%%%%%%%%%%%%%%%
\subsection{Procedure}
%%%%%%%%%%%%%%%%%%%%%%%%%%%%%%
All modeling runs were carried out with time-ordered data, so that period and
epochs were adjustable parameters. Initial periods and epochs were determined
from the spectroscopy, as were the spectral types of the components (except for
the undetectable secondary star of SV~Cam, as we note in the previous section). 
Typically, the following parameters were adjusted: semi-major axis ($a$),
systemic radial velocity ($V_\gamma$), inclination ($i$), temperature of the
secondary star, the star at inferior conjunction at the designated primary
minimum, ($T_2$), the modified Kopal potentials of each star ($\Omega_{1,2}$),
the mass ratio ($q$~=~$M_2/M_1$), the passband luminosity in units of 4$\pi$
($L_{1}$) and, as already mentioned, the epoch ($T_0$) and period ($P$).
Additional suites of parameter adjustments were carried out for BS~Dra and HP~Dra.
Convective atmosphere values for gravity brightening and albedo coefficients
$g$~=~0.32 and $A$~=~0.500, respectively, were assumed for stars similar to 
or cooler than the Sun, and radiative atmosphere values of 1 for both $g$ and $A$ were
assumed for higher temperature stars. A full-precision grid number of 30 was
used for all stars.
%%%%%%%%%%%%%%%%%%%%%%%%%%%%%%%%%%%
\subsection{Solutions}
%%%%%%%%%%%%%%%%%%%%%%%%%%%%%%%%%%%
\subsubsection{SV Camelopardalis}
%%%%%%%%%%%%%%%%%%%%%%%%%%%%%%%%%%%
%%%%%%%%%%%%%%%%%%%%%%%%%%% FIGURES 2 & 3 %%%%%%%%%%%%%%%%%%%%%%%%%%%%%%%%%%%%%%
\begin{figure}[t]
\centerline{\psfig{file=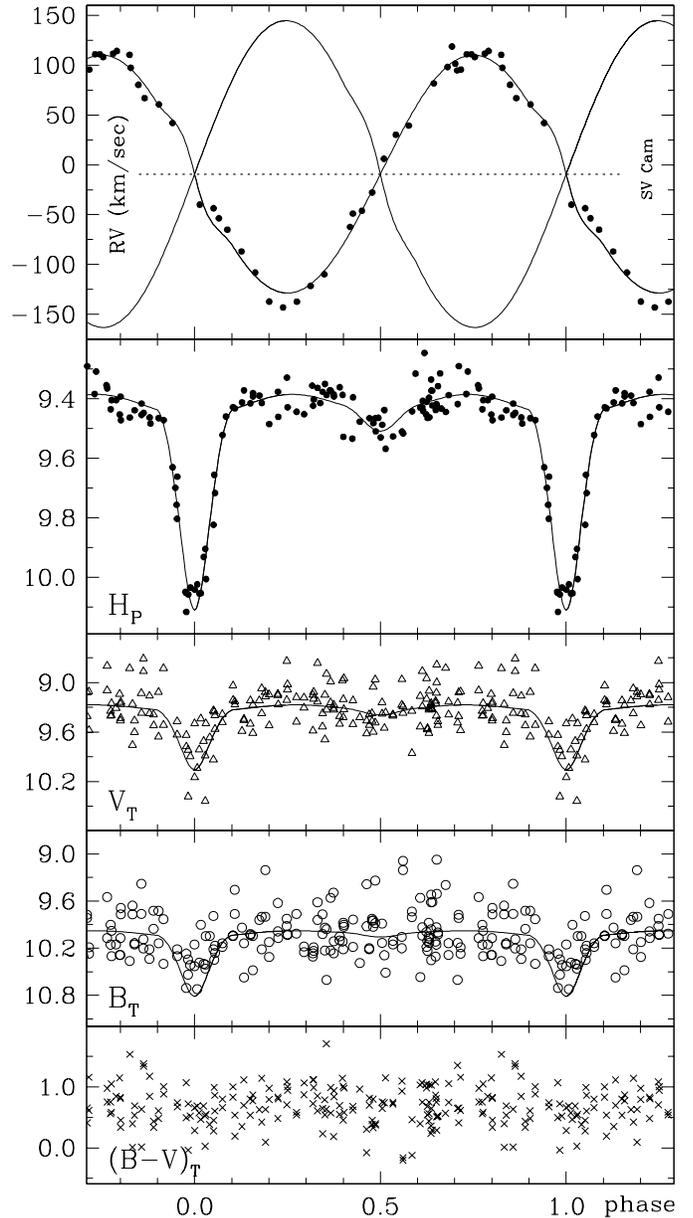,width=8.9cm}}
\caption[]{Hipparcos $H_{\rm P}$ and Tycho $V_{\rm T}$, $B_{\rm T}$, $(B-V)_{\rm T}$ lightcurves and
           the radial velocity curve of SV~Cam folded onto a period $P$~$\sim$~0.593 days. 
           The lines represent the solution given in Table~4.}
\end{figure}
\begin{figure}
\centerline{\psfig{file=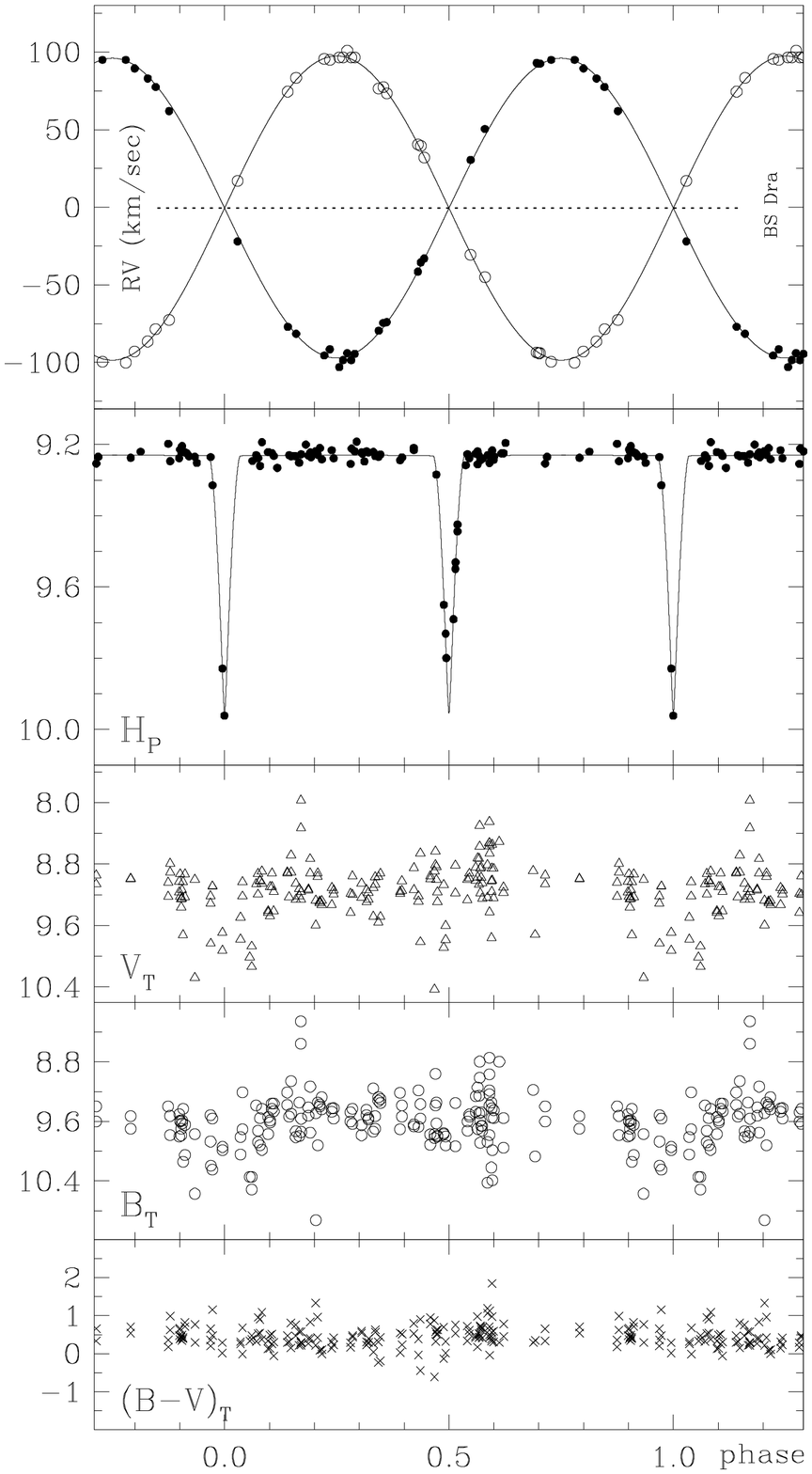,width=8.9cm}}
\caption[]{Hipparcos $H_{\rm P}$ and Tycho $V_{\rm T}$, $B_{\rm T}$, $(B-V)_{\rm T}$ lightcurves of 
           BS~Dra folded onto a period $P$~=~3.364 days. The lines represent the solution 
           given in Table~4.}
\end{figure}
%%%%%%%%%%%%%%%%%%%%%%%%%%%%%%%%%%%%%%%%%%%%%%%%%%%%%%%%%%%%%%%%%%%%%%%%%%%%%%%%%%
The temperature adopted for the hotter star, $T_1$~=~5848~K, was based on a
compromise. The spectra in the Ca~II triplet region indicates a temperature
$\sim$~5800~K or cooler.  The Tycho colors, however, support an slightly
earlier spectral type. The star is rapidly rotating, evolved and sports
numerous surface spots (cf Kjurkchieva et al. 2002 and references therein)
that are responsible for a significant uncertainty when comparing with
template field stars that usually do not have spotty surfaces and 
rotate more slowly. The adopted 5848~K corresponds to a spectral type of G2-G3, as
suggested also by Rainger et al. (1991). The temperature of the secondary
component is determined solely from the light curve analysis.

The single radial velocity curve and the Hipparcos light curve provide the bulk
of the modeling input information because of large scatter in the Tycho light
curves. The analysis was performed fully on two different models. The first
model relies on only the radial velocity and $H_{\rm P}$ measurements, while
the second includes in addition Tycho $B_{\rm T}$ and $V_{\rm T}$ data. Although the
formal fitting error is smaller for the first model, the determination of the
mass ratio in this case depends strongly on the curvature of the light curve
outside of eclipse; therefore, notwithstanding the greater scatter in the Tycho
data, three light curves may provide a better determination of this critical
quantity than one, with appropriate curve weights relating to the intrinsic
scatter in the light curves.

The absolute parameters and other derived quantities for the adopted second
model are given in Table~4 and the fittings are shown in Figure~2. SV Cam is 
indeed a challenging single-lined
spectroscopic binary, nevertheless we obtained a reasonable solution with
accuracies of $\sim$13\% in the masses and $\sim$10\% in the radii. The
cooler star contributes little to $B$ and $V$ light curves and even in the
Ca~II triplet region contributes less than 15\%, according to our model,
which indicates that it is very close to its Roche lobe.

The photometric solution of Albayrak et al. (2001) rests on a temperature of
6440~K for the hotter component and sets of 242 observations each for
$BVR$ passbands.  While their period and orbital separation are quite close
to ours in Table~4 (they quote $P$=0.593071 days and
$a$=3.87$\pm$0.07~$R_{\odot}$), their mass ratio ($q$=0.56) and inclination
($i$=89.6$\pm$0.8) are quite inconsistent with ours. 
There are many possible reasons for the differences:  inclusion or
exclusion of RV data, different light curve data, different limb-darkening
coefficients, and the need for treatment of asymmetries.
The Hipparcos/Tycho light curves present no explicit evidence of
asymmetries that would justify spot modeling, which was therefore not included
in our analyses, 
%A reason for the
%differences is the explicit inclusion of spots in the modeling by Albayrak
%et al. (2001) allowed by their data. The sparse in time and low in number
%Hipparcos/Tycho epoch data that we have adopted in this investigation did
%not allowed detection and therefore modeling of surface spots, 
highlighting
a possible shortcoming in Gaia studies of active surface, single lined
eclipsing binaries.

Our assessments of the SV~Cam system are based exclusively on the treatment
of our adopted input data.  As our anonymous referee has pointed out, these
assessments are not necessarily correct.  Indeed, higher quality
ground-based
data on SV Cam and higher resolution spectroscopy has already indicated that
the secondary component has been observed and that the mass ratio is very
likely 0.6 (somewhat different from our determination, described below).
Although neither the precision in the Hipparcos-Tycho photometry nor the
spectral resolution in our RV data match those of other studies, our
modeling
tools permit distortion in both curves to be treated adequately, given any
discernible spot regions. Therefore our failure to find spot regions may be
an indication of a relatively "quiet" interval in the system's behaviour,
when the activity was below the detection level of the photometry. In such a
case, a mismatch may arise because of differences in activity level at the
separate epochs at which the data were acquired.
The variability impacts our conclusions, and divergence from previous
ground-based investigations indicates that the results for SV Cam are very
important for GAIA mission planning.

%%%%%%%%%%%%%%%%%%%%%%%%%%%%%%%
\subsubsection{BS Draconis}
%%%%%%%%%%%%%%%%%%%%%%%%%%%%%%%
The inspection of the spectra of BS~Dra suggests an F3 spectral type and a
metallicity lower than solar. The temperature of the primary was fixed to
6619~K (Popper 1980) and the modeling was carried out with [Fe/H]=0.0,
$-$0.5, and $-$1.0 metallicities. Although scarcely significant, the fitting
was seen to be slightly better for the $-$0.5 metallicity, as expected for
its proximity with the [Fe/H]=$-$0.4 estimated from our spectra.

The Tycho $B_{\rm T}$ and $V_{\rm T}$ light curves are so noisy that they
essentially add nothing to the modeling, and were excluded from the
analysis. Dropping the information on color variation does not affect the
accuracy of the solution given the almost identical temperature for the two
components. Moreover, the additional information about the curvature at the light curve
maxima was not important to the determination of the mass ratio, unlike the 
case for SV~Cam. It was found that the logarithmic form of the
limb-darkening appeared to improve the fits slightly, and these were
retained for the analyses.

There are very few Hipparcos/Tycho photometric points covering the eclipse
phases. One of the eclipses in the $H_{\rm P}$ light curve has no data points
within 10\% of minimum value and the other has only three points
altogether, all of which are on the descending branch. Yet, if we assume
symmetry, the shape and the depths are defined, even if not with great
precision.  The accuracy of the Hipparcos data for this star
($\sigma_{H_{\rm P}}$=0.016) is high enough to permit the few points in the
light curve minima to be highly weighted to insure that the computed light
curve fits them closely.  All other points were individually weighted
inversely to their standard errors, and scaled. The determination of the
modified Kopal potentials and thus the radii depend critically on this 
close-fitting process.

Both the mass ratio and temperature ratios are near 1.  During modeling, both
the mass ratio and the temperature ratio switched repeatedly across this
boundary. Solutions with either star as the primary component produce nearly
the same fit error, but the adopted converged solution has the epoch given
in the Hipparcos catalogue.  This epoch is, however, half a cycle different
from that of Popper (1971). The alternate model, where the components are
interchanged, converged only with the assumption of a small eccentricity,
but under this circumstance, the argument of periastron, $\omega$, could
not be determined.  The stellar and system parameters are given in Table~4
for the adopted model with [Fe/H]~=~$-$0.5.  Our results can be compared with
those of Russo et al. (1981), who used an earlier version of the
Wilson-Devinney program to analyze previously published $BV$ observations,
probably those of G\"{u}d\"{u}r et al. (1979). Their values for period,
orbital separation, inclination and modified Kopal potentials are close
(even if not always within 1$\sigma$ level) to our results.

%%%%%%%%%%%%%%%%%%%%%%%%%%%%%%%%%
\subsubsection{HP Draconis}
%%%%%%%%%%%%%%%%%%%%%%%%%%%%%%%%%

%%%%%%%%%%%%%%%%%%%%%%%%%%%%%%%%%%%%%%
\begin{figure}[t]
\centerline{\psfig{file=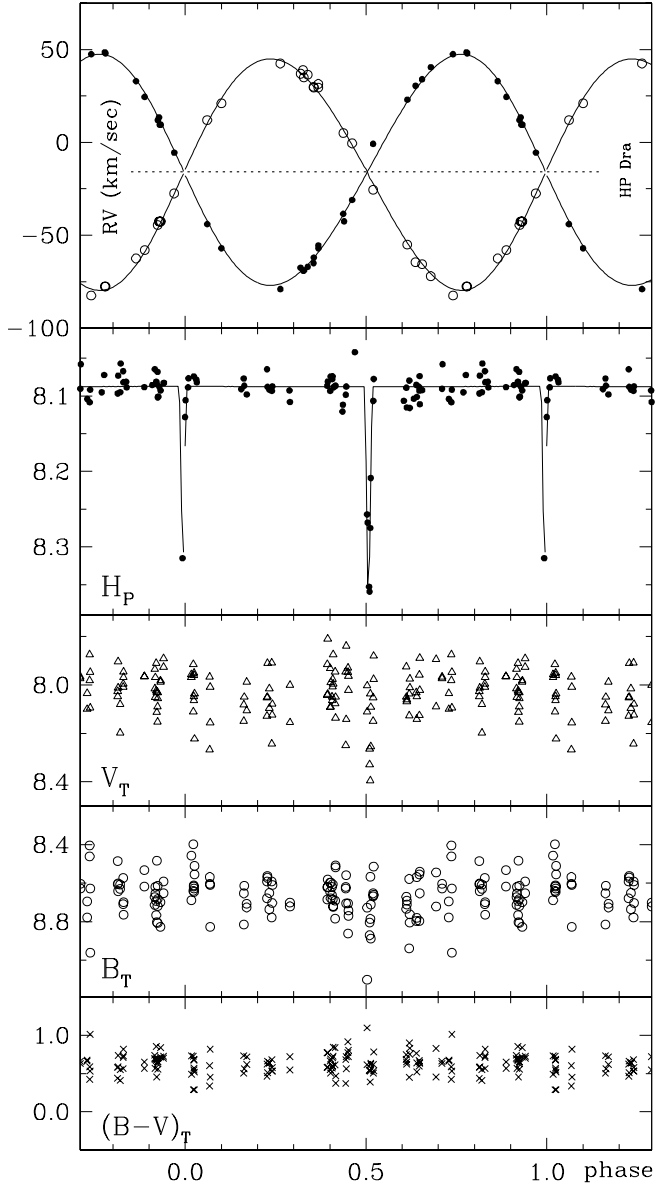,width=8.9cm}}
\caption[]{Hipparcos $H_{\rm P}$ and Tycho $V_{\rm T}$, $B_{\rm T}$, $(B-V)_{\rm T}$ lightcurves of 
           HP~Dra folded onto a period $P$~=~10.761 days. The lines represent the solution 
           given in Table~4.}
\end{figure}
%%%%%%%%%%%%%%%%%%%%%%%%%%%%%%%%%%%%%%%%
HP Dra revealed itself as the most challenging of the 12 objects in this
series of papers. The HP~Dra components appear to be similar in temperature
and luminosity. Because of their great scatter and sparseness at minimum
light, we removed the Tycho data from the modeling sets, even though the
minima are weakly discernible, and modeled the Hipparcos and radial velocity
data alone. There are even fewer data points in the minima for HP Dra than
there are for BS Dra (9 vs. 11). The placements of the data points at
minimum light are critical to the determination of the radii and potentials
of the stars. Although the precision of the Hipparcos data is good
($\sigma_{H_{\rm P}}$=0.012), it still proved very difficult to achieve a
final, fully convergent solution. The major trouble is that the non-coeval 
light and radial velocity curves (about 10 years elapsed between Hipparcos and Asiago
observations) appear to be not fully compatible. If this problem is indeed connected to 
the non-coeval data sets, this would not be a problem for Gaia where all
astrometric, photometric and spectroscopic observations will strictly share
the same epochs. Several strategies were attempted to deal with this circumstance.

\begin{enumerate}
\item Initially, the radial velocities were modeled independently of the
photometry, with assumed temperatures of $\sim$~6000~K for each component,
appropriate for the spectral type ($\sim$~F9) and with equal luminosities. 
This yielded elements that produced an optimal fit to the radial velocity data,
although the assumptions about some of the adopted parameters (such as
$\Omega_{1,2}$) could not be rigorously checked.  The RV modeling indicated
the need to adjust the eccentricity and argument of periastron.  A slight
difference in mass between the components was also found. The parameters which
were adjusted in these trials were: $e$, $\omega$, $V_{\gamma}$, $q$, $T_0$,
and $P$, which are, respectively, the eccentricity, argument of periastron,
system radial velocity, mass ratio ($M_2/M_1$), epoch, and period.
                                                                                                              
\item Further attempts at modeling the combined light and RV curves with these
parameters fixed were also attempted. The results gave minima
which were systematically too shallow. Attempts to weight points in the minima
more heavily also failed to produce convergence.

\item Adjustment of the third light, which failed to improve the
fitting errors.

\item The phased $H_{\rm P}$ light curve data appear to be modulated by a
rough sinusoid at maximum light.  This effect is also seen in the 1989-1990 $V$ light
curve of Kurpinska-Winiarska et al. (2000). Their light curve shows relatively little 
scatter when phased with a similar but independently determined
period. One possibility is a reflection effect of some sort. This is not
likely, given the scale of the orbit and similarity of the temperatures of
the component stars, but it was investigated anyway.  Introduction of a more
complicated reflection effect option in the Wilson-Devinney program also
failed to provide adequate fittings to the $H_{\rm P}$ light curve, even
when the size of the primary star was initially increased (by decreasing its
potential).  The computing time increases greatly when this option is
adopted in treating an eccentric orbit case, such as HP~Dra, so we performed only
one trial with the enhanced (2-pass) reflection effect.  The fitting was not
quite as good as for the simple reflection case, in which the illuminating
star is assumed to be a point source and only one reflection is treated.
                                                                                                              
\item Because intrinsic variation may also be affecting the light curve, we
attempted to introduce a large starspot onto the visible hemisphere of the
component at inferior conjunction during eclipses: either a hot spot on the
'secondary' star visible at primary minimum (phase 0.0) or a cool, spot on
the 'primary' star visible at secondary minimum (phase 0.5).  With spots
centered on a longitude of 180$\degr$ and with a radius of 90$\degr$, the
modulation was adequately fit but a solution with this artifice still failed
to converge, in either spot placement case.  The existence of a spot on
stars just slightly hotter than the Sun would not be surprising, but we have
not found lower fit errors by assuming convective envelopes in these
components (although not every model was tested), thus star spots appear to
be both unnecessary and insufficient to model this system, given only the
current suite of data that have been modeled.

\item An investigation of variability in the residuals failed to show
periodicities differing slightly from the derived eclipsing period, viz., over
the range 3.364000000 to 3.364033141 days.

\item Adjustment of d$P$/dt equally failed to produce an improvement.
                                                                                                              
\item On the contrary, adjustment of the d$\omega$/dt term did produce an
improvement in fitting of both light curve and radial velocity curves (when
modeled simultaneously). This finally yielded a converged solution that
allowed us to combine into a coherent picture both photometric data and radial
velocities in spite of the 10 year elapsed between them. This is the
adopted solution presented in Table~4.
\end{enumerate}                                                                                                              
                                                                                                              
Even though the system seems to be at or just beyond the data-paucity limit
for analysis of partially eclipsing systems, we were able to achieve a fully
convergent and self-consistent solution. It is worth repeating here that the two
major problems encountered in the analysis of HP~Dra (i.e., non-coevality of
photometric and radial velocity curves and paucity of the photometric points
covering the eclipse phases) will not pertain to Gaia (cf. Katz et al.
2004). The satellite will obtain simultaneous spectroscopic and
photometric data during the planned 5 year mission, the photometry at about
$\sim$355 independent epochs (cf. Munari et al. 2004) compared to the
$\sim$100 Hipparcos epochs.

%%%%%%%%%%%%%%%%%%%%%%%%%%%%%%%%%%%%%%%%%%%%%%%%%%%%%%%%%%%%%%%%%%%%%%%%%%%%%%%%%
% TABLE 5
%%%%%%%%%%%%%%%%%%%%%%%%%%%%%%%%%%%%%%%%%%%%%%%%%%%%%%%%%%%%%%%%%%%%%%%%%%%%%%%%%
\begin{table}[!b]
\begin{center}
\tabcolsep 0.08truecm
\caption{Comparison between the Hipparcos distances and those derived from the 
parameters of the modeling solution in Table~4 assuming an uncertainty of $\pm$150~K
in the temperature adopted for the primary in each system.}
\begin{tabular}{lcc}
\hline
          &Hipparcos             & this paper      \\
          & (pc)                 &  (pc)           \\
          &                      &                 \\
SV~Cam    & $85^{93}_{79}$       &  $87\pm{8}$     \\ 
          &                      &                 \\
BS~Dra    & $208^{246}_{181}$    &  $172\pm{8}$    \\
          &                      &                 \\
HP~Dra    & $80^{85}_{76}$       &  $73\pm{4}$     \\
          &                      &                 \\
\hline
\end{tabular}
\end{center}
\end{table}
                                                                                                              
Nevertheless, the resulting model cannot be
said to be a truly accurate one.  With only three points in one of the minima
(and only one point near minimum value), it is clear that the relative surface
brightnesses of the two stars cannot be obtained from the Hipparchos photometry
alone, the depth providing less information than the branches of the minima,
which the model must fit. In fact the light curve produced by
Kurpinska-Winiarska et al. (2000) shows the secondary minimum to be the
shallower. This suggests that there is indeed a temperature difference between
the components, but in the opposite sense from that found here. The large
uncertainty in the temperature difference is therefore reasonable.  The
difference in the potentials is determined largely by the phases of contact and
is therefore also uncertain, but the outer contacts are not badly defined by
the data. Our model indicates that there is a small orbital eccentricity, in
rough agreement with a preliminary value ($e$=0.043) found by
Kurpinska-Winiarska et al. (2000). As far as we know, there has been no
evidence presented for a third body in the system, but we could not achieve
convergence in the modeled data set without adjustments of the apsidal motion.
For HP~Dra, complete light- and radial velocity curves exist and the full
analysis promised by Kurpinska-Winiarska et al. (2000) is awaited with great
interest. If the apsidal motion mentioned here can be confirmed, additional
photometry across a range of bandpasses, times of minima, and spectroscopy,
should be pursued.

It is possible that further spot modeling on a fuller data set will be able
to resolve the question of the true nature of HP Dra.  For now, we indicate
only the possibility of apsidal motion, and the limitations for GAIA models
given our current suite of tools to test that mission's capability.

%%%%%%%%%%%%%%%%%%%%%%%%%%%%%%%%%%%%%
\section{Conclusions}
%%%%%%%%%%%%%%%%%%%%%%%%%%%%%%%%%%%%%
The distances derived from the orbital solutions are compared to those from
the Hipparcos Catalogue in Table~5.  Despite the fairly large uncertainties
in the fundamental parameters, the derived distances to all three systems
agree with the astrometric distances from Hipparcos-Tycho within or very
close to their one-sigma errors. 

The overall agreement confirms similar findings in previous papers
in this series and together they reinforce confidence in the overall
quality and astrophysical potential of orbital modeling of eclipsing binary
data to be obtained by Gaia.

As in other papers in this series, we have been discriminate in the use of the
Tycho mission data, because of its limited precision, and effectively used them
in the present paper only for the SV Cam system for reasons mentioned earlier.
For fainter systems, the Tycho data are useful only for the limited mean color
information that they provide.  For the BS Dra and HP Dra systems, they proved
unsuitable for modeling purposes.

Although we conclude that the results for the three systems presented here are
plausible, in context with existing, higher quality data other than our
adopted data suite, we note the possibility of systematic error in
the analyses based on photometry only from Hipparcos and Tycho missions, even
when these are coupled with spectroscopy that closely matches GAIA's. In
particular, the results for SV Cam, and possibly also for HP Dra, point to the
need to diagnose spot regions in systems detected by GAIA. Of course,
it is possible that a system will appear "quiet" during times of data
acquisition, but GAIA will acquire all its types of data at the same instant,
so they will be self-consistent, at least. Moreover,
the suite of photometric passbands recently selected for
GAIA permits the discrimination of system properties from photometric indices
of passbands across a broad region of the spectrum, making the color effects
of spotted regions easier to detect.

%%%%%%%%%%%%%%%%%%%%%%%%%%%%%%%%%%%%%%%%%%%%%%%
\begin{acknowledgements}
%%%%%%%%%%%%%%%%%%%%%%%%%%%%%%%%%%%%%%%%%%%%%%%
It is a pleasure to acknowledge the help of W. Van Hamme, C. R. Stagg and Dirk 
Terrell, for limb-darkening coefficients for the Gaia-RVS, Hipparcos, and Tycho
passbands, flux ratio files creation, and desk-top limb-darkening 
interpolation software, respectively.  
The authors are grateful to our anonymous referee for several insightful
comments and suggestions that have been incorporated in the paper.
This work was supported in part by Italian 
MIUR COFIN 2001 and ASI 20001 grants, and by Canadian NSERC grants.
\end{acknowledgements}

%%%%%%%%%%%%%%%%%%%%%%%%%%%%%%%%

%\clearpage
\end{document}